\documentclass[prd,11pt,showpacc]{revtex4}
\usepackage{graphics}
\usepackage{epsfig}
\def\bra{\langle }
\def\ket{\rangle }
\usepackage{amsmath,amssymb,amsfonts,textcomp}
\usepackage{mathrsfs}
\usepackage{eucal}
\usepackage{amssymb}
\newcommand{\eq}[1]{Eq.~\eqref{eq:#1}}

\def\dpdf#1{F_{#1}}
\def\spdf#1{q_{#1}}

\newcommand{\nn}{\nonumber}
\newcommand{\df}{\mathrm{d}}
\newcommand{\img}{\mathrm{i}}

\newcommand{\si}{\sigma}

\newcommand{\up}{\uparrow}
\newcommand{\down}{\downarrow}

\newcommand{\lqcd}{\Lambda_\mathrm{QCD}}

\begin{document}
\title{Double Parton Correlations
in Constituent Quark Models }

\title{Double parton correlations in constituent quark models}

\author{M. Rinaldi$^1$\footnote{E-mail address: matteo.rinaldi@pg.infn.it},
S. Scopetta$^{1}$\footnote{
E-mail address: sergio.scopetta@pg.infn.it},
V. Vento$^{2}$\footnote{E-mail
address: vicente.vento@uv.es}
}
\affiliation
{\it
(1)
Dipartimento di Fisica, Universit\`a degli Studi di Perugia, and
INFN, sezione di Perugia, via A. Pascoli
06100 Perugia, Italy;
\\
(2)
Departament de Fisica Te\`orica, Universitat de Val\`encia
and Institut de Fisica Corpuscular, Consejo Superior de Investigaciones
Cient\'{\i}ficas, 46100 Burjassot (Val\`encia), Spain.
}

%\date{}

\begin{abstract}
Double parton correlations, 
having effects on the double parton
scattering processes occurring in high-energy hadron-hadron collisions,
for example at the LHC, are studied
in the valence quark region,
{  by means of} constituent quark models.
In this framework, two particle correlations are present 
without any additional prescription, at variance 
with what happens, for example, in independent particle models, 
such as the MIT bag model in its simplest version.
From the present analysis, 
conclusions similar to the ones obtained recently in a modified version of the 
bag model can be drawn: 
correlations in the longitudinal momenta of the active quarks are found
to be sizable, while those in transverse momentum are much smaller.
However, the used framework allows to understand clearly
the dynamical origin of the correlations. In particular, it is
shown that the small size of the correlations in transverse momentum
is a model dependent result, which would not occur if models
with sizable quark orbital angular momentum were used to describe 
the proton. Our analysis permits therefore to clarify the dynamical origin
of the double parton correlations and to
establish which, among the features
of the results, are model independent. 
The possibility to test experimentally the studied effects is 
discussed.
\end{abstract}
   

\maketitle

%\onecolumn
\section{Introduction}

Long time ago, the relevance of
multiple hard partonic collisions taking place 
within a single hadronic scattering, 
the so called multiple parton interactions (MPI),
has been addressed and studied ~\cite{Paver:1982yp}.
Although MPI are higher twist, i.e.~they are suppressed by a power of 
$\lqcd^2/Q^2$ with respect to single parton
interactions, $Q$ {  being} the partonic center-of-mass energy 
in the collision,
experimental evidence of these kind of processes
has been obtained, already some years ago \cite{livio}.
MPI are of great relevance for LHC Physics, where they represent
a background for the search of new Physics. 
It is not surprising therefore that
a strong debate around MPI has been arising in recent years, when
several dedicated Workshops have been organized, starting from that 
illustrated in Ref. \cite{pg}. 
In this work we concentrate on double parton scattering (DPS) 
processes.
It is now understood that DPS 
contributes, e.g., to same-sign $WW$ and same-sign 
dilepton production
~\cite{Kulesza:1999zh,Cattaruzza:2005nu,Maina:2009sj,Gaunt:2010pi} 
and represents a background for Higgs studies in the channel 
$pp \to WH \to \ell \nu b\bar b$ 
\cite{DelFabbro:1999tf,Hussein:2006xr,Bandurin:2010gn,Berger:2011ep}. 
Evidence of the occurrence
of DPS at LHC has been already established~\cite{Aad:2013bjm}.
Two comprehensive papers reporting the status of the art {  of}
theoretical knowledge {  for} these processes have recently appeared
\cite{Diehl:2011yj,Manohar:2012jr}.
New signatures of DPS producing double Drell-Yan processes
have been studied recently
~\cite{Kasemets:2012pr}.
 The dynamical origin of double parton correlations,
having potential effects in DPS,
in semi-inclusive deep inelastic scattering and in hard
exclusive processes, has been the
subject of recent comprehensive studies \cite{weiss}.

In the original work~\cite{Paver:1982yp}, 
the DPS cross section was written in terms
of double parton distribution functions (dPDF) 
$\dpdf{ij}(x_1,x_2,\vec z_\perp)$, describing the joint probability 
of having two partons with flavors $i,j=q, \bar q,g$, 
longitudinal momentum fractions $x_1,x_2$ and transverse separation 
$\vec z_\perp$ inside a hadron:
%%%
\begin{align}  \label{eq:si_old}
  \df \si &= \frac{1}{S} \sum_{i,j,k,l} \int\! \df^2 \vec z_\perp\, 
\dpdf{ij}(x_1,x_2,\vec z_\perp,\mu) 
\dpdf{kl}(x_3,x_4,\vec z_\perp,\mu)  \nn \\ & \quad \times 
  \hat \si_{ik}(x_1 x_3 \sqrt{s},\mu) \hat \si_{jl}(x_2 x_4 \sqrt{s},\mu)
\,.\end{align}
%%% 
The partonic cross sections $\hat \si$ represent the hard,
short-distance processes, while
$S$ is a symmetry factor arising if identical particles are present in 
the final state. In \eq{si_old}, contributions due 
to diparton flavor, spin and color correlations are neglected, 
as well as parton-exchange interference 
contributions~\cite{Diehl:2011yj,Manohar:2012jr,Mekhfi:1985dv,Diehl:2011tt}.
These correlations are present in QCD and could be phenomenologically
relevant.
{  Positivity bounds on dPDFs have been recently derived
for polarized dPDFs \cite{oggi}}.

Two main assumptions are usually made in DPS analyses: 
\\
i)
the dependence on the 
transverse separation and on the momentum fractions 
or parton flavors are not correlated:
%%%
\begin{equation} \label{eq:zfact}
  \dpdf{ij}(x_1,x_2,\vec z_\perp,\mu) = \dpdf{ij}(x_1,x_2,\mu) 
T(\vec z_\perp,\mu)
\,;\end{equation}
%%%
ii)
a factorized form is chosen also for the $x_1,x_2$ dependence:
%%%
\begin{equation} \label{eq:xfact}
\begin{split}
  &\dpdf{ij}(x_1,x_2,\mu) \\
   &= \spdf{i}(x_1,\mu) \spdf{j}(x_2,\mu)\, \theta(1-x_1-x_2) (1-x_1-x_2)^n
\,,
\end{split}
\end{equation}
%%%
where $q$ is the usual parton distribution function (PDF).  
$\theta(1-x_1-x_2) (1-x_1-x_2)^n$ introduces the 
kinematic constraint $x_1+x_2 \leq 1$, and $n>0$
is a parameter to be fixed phenomenologically. 
dPDFs, describing soft {  physics},
are nonperturbative functions. 

dPDFs cannot be evaluated in QCD but, as it happens for the usual
PDFs, they can be at least estimated  
at a low scale, 
$Q_0 \sim \lqcd$, the so called hadronic scale, using quark models.
{In order to compare them with data taken at a momentum scale
$Q>Q_0$, the results of model calculations
should be evolved using perturbative QCD (pQCD).
This evolution of dPDFs, 
namely the way they change from  
$Q_0$ to $Q>Q_0$, is  currently a matter of debate
~\cite{Kirschner:1979im,Shelest:1982dg,Diehl:2011yj,Manohar:2012jr,
Diehl:2011tt,
agg1,agg2,agg3,agg4,Manohar:DPS2}.}
The resolution of this debate will be instrumental to relate 
not only data from different experiments  but also model 
calculations at the hadronic scale to LHC data.  
In this way, the analysis of data
involving DPS can be guided.

A first  {  model} calculation is the one recently
presented in Ref.~\cite{man_bag}, where
a bag model framework for the proton~\cite{Chodos:1974pn} has been
considered to evaluate dPDFs in the valence region,
{  at the hadronic scale $Q_0$, without evolution 
to $Q>Q_0$.}
As in any model where
the valence quarks carry all the momentum, in the bag model
{  the scale $Q_0$} has to be taken quite low.
If the bag were assumed to be rigid,
as in early calculations, 
corresponding to the so-called cavity approximation
~\cite{Jaffe:1974nj},
the quarks would be independent and none of the relevant
correlations described by dPDFs would be found.
In Ref.~\cite{man_bag}, therefore, a prescription to recover
momentum conservation, leading to quark correlations in the bag,
already applied in model calculations of parton distributions,
has been used ~\cite{Benesh:1987ie,Schreiber:1991tc,Wang:1982tz}.
The main outcome of Ref.~\cite{man_bag} is that, in the
modified MIT bag model,
the approximation \eq{zfact} holds reasonably well, 
but \eq{xfact} is strongly violated. 
{ Problems with \eq{xfact} had already been pointed out 
in Ref.~\cite{snig1,Gaunt:2009re,snig2}.} 

The analysis of Ref.~\cite{man_bag} is retaken here in a constituent
quark model (CQM) framework. 
CQM calculations of parton distributions have been proven
to be able to predict the gross features of PDFs \cite{trvlc,h1},
generalized parton distributions (GPDs) \cite{vlc_pg}
and transverse momentum dependent parton distributions (TMDs) 
\cite{siv_bm}. Similar expectations motivated the present analysis.
With respect to the approach of Ref.~\cite{man_bag},
the non relativistic (NR) dynamics used here includes from the very beginning
correlations into the scheme, so that, for example,
the dynamical origin of the breaking of the approximation
\eq{zfact} is clarified.
As will be seen, this analysis helps 
to understand which among the conclusions
of Ref.~\cite{man_bag} are model dependent.
It is important to stress from the very beginning that
this kind of analysis, as the one in Ref. \cite{man_bag},
is valid at the low hadronic scale corresponding to the model
and only in the valence quark region. For this study to be
directly used in the LHC data analysis, having for the moment high
statistics only for very small values of $x$, far 
therefore from the valence region,
and corresponding to high momentum transfer,
additional studies, in particular the QCD evolution
of the model results, are necessary.
This investigation, like the one in Ref. \cite{man_bag}, is therefore
a first exploratory estimate, thought to understand the dynamical
origin of the different momentum correlations and to guide
experimental measurements accordingly. 

The paper is structured as follows.
In the next section, the formalism to evaluate
the dPDFs in a CQM is clarified and the main equations
presented. In the third one, results are illustrated and discussed.
Eventually, conclusions are drawn in the last section.

\section{Constituent quark model description of double parton distribution
functions}

As already stated in the Introduction,
we are going to use a CQM framework and therefore only
correlations for two valence quarks of 
flavor {  $i,j$} can be studied.
Let us work in momentum space,
in terms of the Fourier-transformed 
distribution $\dpdf{ij}(x_1,x_2,{\vec k}_\perp)$, defined as follows:
%%%
\begin{eqnarray} 
\label{ft}
\dpdf{ij}(x_1,x_2,{\vec k}_\perp) 
&= \int\! d \vec z_\perp \, e^{\img \vec z_\perp \cdot \vec k_\perp} 
\dpdf{ij}(x_1,x_2,{\vec z}_\perp)~.
\end{eqnarray}
The formalism introduced in Ref.~\cite{Manohar:2012jr} will be used.
Color-correlated and interference dPDFs will not be considered,
being Sudakov suppressed at high energies
\cite{Mekhfi:1988kj,Manohar:2012jr}. 

{  In a NR quark model, the color-summed dPDF for two valence quarks,
one with flavor $q_1$ and longitudinal momentum $x_1$, 
the other with flavor $q_2$ and longitudinal momentum $x_2$,
with polarization $s_1,s_2$ in the proton, respectively, is:}
\begin{eqnarray}
\label{uno}
F_{q_1q_2}(x_1,x_2,\vec k_\perp) & = & 
\int {d \vec k_1 } 
{d \vec k_2 } 
\int {d \vec k_1' } 
{d \vec k_2' } 
\, n_{q_1q_2}(\vec k_1,s_1;\vec k_2,s_2;\vec k_1',s_1;\vec k_2',s_2)
\nonumber
\\
& \times &
\delta \left ( \vec k_1' - \vec k_1 - {\vec k_\perp} \right )
\delta \left ( \vec k_2' - \vec k_2 + {\vec k_\perp} \right )
\delta \left ( x_1 - {k_1^- \over P^-} \right )
\delta \left ( x_2 - {k_2^- \over P^-} \right )~,
\end{eqnarray}
where light cone components, given by $a^\pm= p_0 \pm p_3$ for
a generic 4-vector $a^\mu$ have been used for the quark
and proton momenta and   
the two quark, spin dependent, off diagonal momentum distribution
\begin{eqnarray}
\label{manybody}
n_{q_1q_2}(\vec k_1,s_1;\vec k_2,s2;\vec k_1',s_1;\vec k_2',s2)
& = &  3 {\sum_{s_3}} 
\int d \vec{k}_3 
\, \Phi^*_{} \left (\vec k_1, s_1; 
\vec k_2, s_2; \vec k_3, s_3 \right ) 
\hat P_{q_1}(1) \hat P_{q_2}(2)
\hat P_{s_1}(1) \hat P_{s_2}(2)
\nonumber
\\
& \times &
\Phi \left (\vec k_1', s_1; 
\vec k_2', s_2; \vec k_3, s_3 \right ) 
\end{eqnarray}
has been introduced, using momentum space 
wave functions.
In Eq. (\ref{manybody}), the flavor and spin projectors,
given by
\begin{eqnarray}
\hat P_{u(d)}(i) = { 1 \pm \tau_3(i) \over 2}
\end{eqnarray}
and
\begin{eqnarray}
\hat P_{\up(\down)}(i) = { 1 \pm \sigma_3(i) \over 2}~,
\end{eqnarray}
respectively,
for the particles 1, 2 appear.
The factor 3 in Eq. (\ref{manybody}) represents
the number of valence quark pairs in the proton.
It arises from the fact that any pair is equivalent to each other,
since the proton ground state wave function is symmetric, 
once the color degrees of freedom
are factorized out.

By taking into account properly the corresponding flavor
and spin projectors, any kind of spin and flavor correlations
can be easily evaluated.
For the sake of definiteness,
in this paper we will concentrate on spin averaged dPDFs,
which we call, for brevity, distributions  $q_1q_2$.
In a CQM framework, using translational invariance, they
have therefore, according to Eqs. (\ref{uno}) and (\ref{manybody}), 
the following form:
\begin{eqnarray}
\label{q1q2}
q_1q_2(x_1,x_2, k_\perp)
& = & 
3 \int {d \vec k_1 } 
{d \vec k_2 }
{d \vec k_3 }
\, \psi^*_{} \left (\vec k_1 + { \vec k_\perp \over 2}, 
\vec k_2 - {\vec k_\perp \over 2}, \vec k_3 \right ) 
\hat P_{q_1}(1) \hat P_{q_2}(2)
\nonumber
\\
& \times & 
\, \psi_{} \left ( \vec k_1 - { \vec k_\perp \over 2},
\vec k_2 + {\vec k_\perp \over 2}, \vec k_3 \right )
\delta \left ( \vec k_1 + \vec k_2 + \vec k_3 \right )
\delta \left ( x_1 - {k_1^- \over P^-} \right )
\delta \left ( x_2 - {k_2^- \over P^-} \right ) 
\nonumber
\\
& = & 
3 \int {d \vec k_1 } 
{d \vec k_2 }
\, \psi^*_{} \left (\vec k_1 + { \vec k_\perp \over 2}, 
\vec k_2 - {\vec k_\perp \over 2} \right ) 
\hat P_{q_1}(1) \hat P_{q_2}(2)
\nonumber
\\
& \times & 
\psi_{} \left ( \vec k_1 - { \vec k_\perp \over 2},
\vec k_2 + {\vec k_\perp \over 2} \right)
\delta \left ( x_1 - {k_1^- \over P^-} \right )
\delta \left ( x_2 - {k_2^- \over P^-} \right )~,
\end{eqnarray}
where the intrinsic wave function $\psi(\vec k_1, \vec k_2)$ has
been introduced and $k_\perp=|\vec k_\perp|$.
A similar expression was presented already in Ref. \cite{Manohar:2012jr},
obtained as the NR limit of the general definition of the dPDFs, given in terms
of light-cone quantized operators and states. 
This equation allows one to evaluate $q_1q_2$ 
in any CQM, by using the corresponding intrinsic momentum space
wave functions.

In our presentation, we will use two different CQM:

%\begin{itemize}

%\item
$i)$ a simple harmonic oscillator (HO) potential model,
assuming an SU(6) spin-flavor structure, with the proton
representing the ground state. In this case,
the intrinsic wave function {  for the nucleon N} can be written:
\begin{equation}
\label{wf}
|N^2 S_{1/2}\ket_S =
\psi_{}^{\cal{S}}(\vec k_1,\vec k_2)
=
{\sqrt{3}^{3/2} \over 
\pi^{3/2} \alpha^3 } e^{-{
(\vec k_1^2 + \vec k_2^2 + \vec k_1 \cdot \vec k_2) \over
{\alpha^2} }} \times SU(6)
\end{equation}
where the spectroscopic notation $|^{2S+1}X_J \rangle_t$,
with $t=A,M,S$ being the symmetry type, has been used and
$SU(6)$ is the usual SU(6) spin-flavor proton wave function
($\chi$ refers to spin, $\Phi$ to flavor):
\begin{equation}
SU(6) = { 1 \over \sqrt {2}} ( \chi_{MS}\Phi_{MS} + \chi_{MA}\Phi_{MA})~,
\label{su6}
\end{equation} 
whose explicit expression is given in textbooks and not repeated here.
The parameter $\alpha^2 = m \omega$ of the HO
potential is fixed to the value 1.35 fm$^{-2}$, in order
to reproduce the slope of the proton charge form factor at zero 
momentum transfer.
Despite {  its} simplicity, it is perfectly clear that
in this model dynamical correlations between the quarks are present
from the very beginning, due to the $\vec k_1 \cdot \vec k_2$
term in the wave function, preventing it from being a simple
product of two HO ground state wave functions.
It is remarkable that this does not happen in the MIT
bag model, whose simplest realization describes just
independent, uncorrelated particles.
In Ref. \cite{man_bag}, correlations are introduced
by properly modifying the simplest bag model scenario.
In a potential model, we have them even in the simplest
case. Therefore it is very appropriate
to study double parton correlations in this framework.

%\item
$ii)$
the CQM of Isgur and Karl (IK) \cite{ik}.
In this model,
the proton wave function is obtained by adding
a one gluon exchange contribution
to a confining HO potential,
along the line addressed already in Ref. \cite{ruju};
including contributions up to the $2 \hbar \omega$ shell,
neglecting a very small ${\cal{D}}$ wave contribution, 
the proton 
$|N \rangle$
is given by the
following admixture of states
\begin{eqnarray}
|N \rangle 
=
\psi_{}(\vec k_1,\vec k_2,s_1,s_2,s_3)
=
a_{\cal {S}} |N^2 S_{1/2}\ket_S +
a_{\cal {S'}} |N^2 S'_{1/2}\ket_S +
a_{\cal M} |N^2 S_{1/2}\ket_M~.
\label{ikwf}
\end{eqnarray}
The coefficients were determined by spectroscopic properties to be
$ a_{\cal {S}} = 0.931$,
$ a_{\cal {S'}} = -0.274$,
$ a_{\cal M} = -0.233$. 
\cite{mmg}.
In this case, the parameter $\alpha^2 = m \omega$ of the HO
potential is fixed to the value 1.23 fm$^{-2}$, in order
to reproduce the slope of the proton charge form factor at zero 
momentum transfer \cite{mmg}.
If $a_{\cal {S}}=1$ and
$ a_{\cal {S'}}= 
a_{\cal M} = 0$ the simple HO model 
is recovered.

The formal expressions of the wave functions appearing in Eq. (\ref{ikwf}),
yielding the IK model, can be found in Ref. \cite{mmg,ik2},
given in terms
of the following sets of
conjugated intrinsic coordinates
\begin{eqnarray}
\vec R = { 1 \over \sqrt{3} } ( \vec r_1 + \vec r_2 + \vec r_3 )
& \leftrightarrow &
\vec K = { 1 \over \sqrt 3} ( \vec k_1 + \vec k_2 + \vec k_3 )
\nonumber ~,
\\
\vec \rho = { 1 \over \sqrt 2} ( \vec r_1  - \vec r_2 )
& \leftrightarrow &
\vec k_{\rho} = { 1 \over \sqrt 2} ( \vec k_1 -  \vec k_2 )
\nonumber ~,
\\
\vec \lambda = { 1 \over \sqrt 6} ( \vec r_1 + \vec r_2 - 2 \vec r_3 )
& \leftrightarrow &
\vec k_{\lambda} = { 1 \over \sqrt 6} ( \vec k_1 + \vec k_2 - 2 \vec k_3 )
~.
\label{coor}
\end{eqnarray}
There are many good reasons to use the IK model to estimate
{  dPDFs}.
First of all, the IK is a typical CQM,
successful in reproducing the low energy properties of the nucleon,
such as the spectrum and the electromagnetic form factors
at small momentum transfer \cite{ik,mmg}.
In studies of DIS phenomenology, IK based model calculations were
able to describe the gross features of PDFs \cite{trvlc,h1},
generalized parton distributions (GPDs) \cite{vlc_pg}
and transverse momentum dependent parton distributions (TMDs) 
\cite{siv_bm}. 
%\end{itemize}

From now on, for the sake of definiteness, {  we will
consider 
$q_1=u_1$, $q_2=u_2$, i.e., the correlations between
a quark $u$
with longitudinal momentum fraction $x_1$ and the other quark $u$
with longitudinal momentum $x_2$.
Since this situation cannot be distinguished from that in which
the momenta are exchanged between the two $u$ quarks, 
we will actually show and discuss our results for the distribution 
\begin{equation} 
{uu}(x_1,x_2, k_\perp) = {u_1u_2}(x_1,x_2, k_\perp)
+ {u_2u_1}(x_1,x_2, k_\perp) = 2{u_1u_2}(x_1,x_2, k_\perp)~.
\label{vera}
\end{equation} 
This quantity is the one usually discussed in the literature,
e.g., in Refs. \cite{man_bag,Gaunt:2009re}, and 
this choice of presentation {  allows} therefore an easy comparison of our
results with those of other authors.
The normalization of the dPDF Eq. (\ref{vera}), built according to
Eqs. (\ref{q1q2}) - (\ref{su6}), is found to be
\begin{equation}
\label{norm}
\int d x_1 d x_2 
{uu}(x_1,x_2, k_\perp=0) = N_u(N_u-1)~,
\end{equation}
$N_u$ being the number of $u$ quarks in the system
under investigation. For the proton discussed here, 
$N_u=N_u(N_u-1)=2$.
In this case, the dPDF represents a number density, its norm
yielding just the number of $u$ quarks, with any momentum,
in the proton.
}

By inserting the wave function Eq. (\ref{wf}) in Eq. (\ref{q1q2}),
one gets, in the HO model:
\begin{equation} 
\label{uuho}
{uu}^{HO}(x_1,x_2,k_\perp) = 
C {{e^{ - { k_\perp^2   \over 2 \alpha^2} }}} 
\int {d \vec k_1 } 
{d \vec k_2 } e^{ - f(\vec k_1, \vec k_2,\alpha)}
\delta \left ( x_1 - {k_1^- \over P^-} \right )
\delta \left ( x_2 - {k_2^- \over P^-} \right )~,
\end{equation}
where $C = 2 (\sqrt{3}/(\pi \alpha^2))^3$ and
$f(\vec k_1, \vec k_2,\alpha)=
2 (k_1^2 + k_2^2 + \vec k_1 \cdot \vec k_2)/\alpha^2$.
The delta function is worked out considering 
the nucleon with mass $M$ at rest and 
the quark with mass $m \simeq M/3$ on mass shell, with energy
$k_0 = \sqrt{m^2 + k^2}$. 
In this way one has, for $i=1,2$:
\begin{eqnarray}
x_i = {k_i^- \over P^-} = {\sqrt{m^2 + k_i^2} - k_{iz} \over M}~.
\end{eqnarray}
From 
this assumption, a
{\small} support violation
arises, e.g., a {  small}  tail of the longitudinal momentum distribution
is found for $x > 1$. 
From Eq. (\ref{uuho}), it is clear that $uu$ depends on 
$k_\perp$,
only in the exponent of the Gaussian function outside the integral.
\begin{figure}[t]
\begin{minipage}[t] {70 mm}
\vspace{7.0cm}
\includegraphics{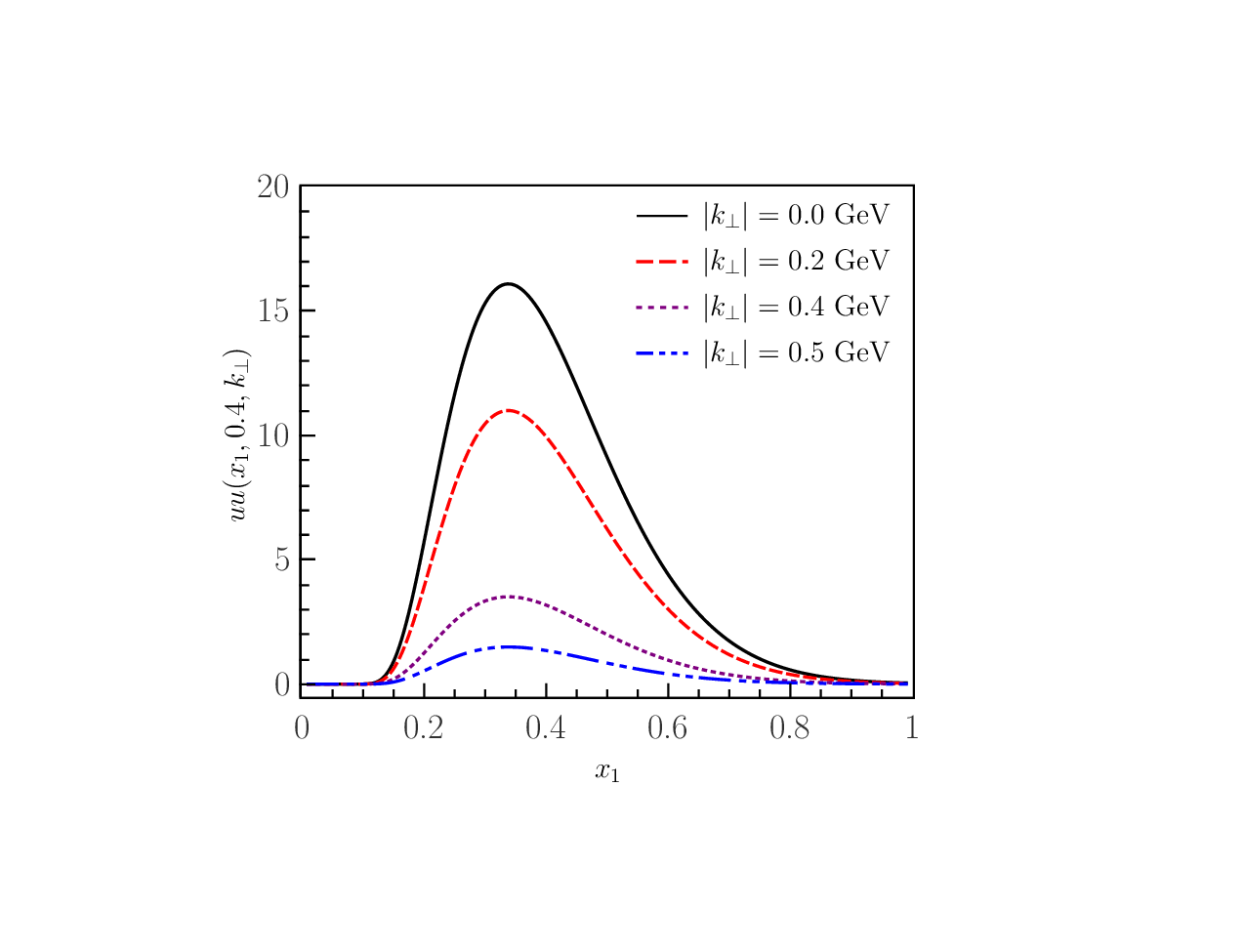}
%\vskip 6.5cm
\caption{The dPDF $uu(x_1,x_2,k_\perp)$ in the HO model,
Eq. (\ref{uuho}), for $x_2=0.4$ at four values of $k_\perp$ 
.}
\thispagestyle{empty}
\end{minipage}
\hspace{\fill}
\begin{minipage}[t] {70 mm}
\vspace{7.0cm}
\includegraphics{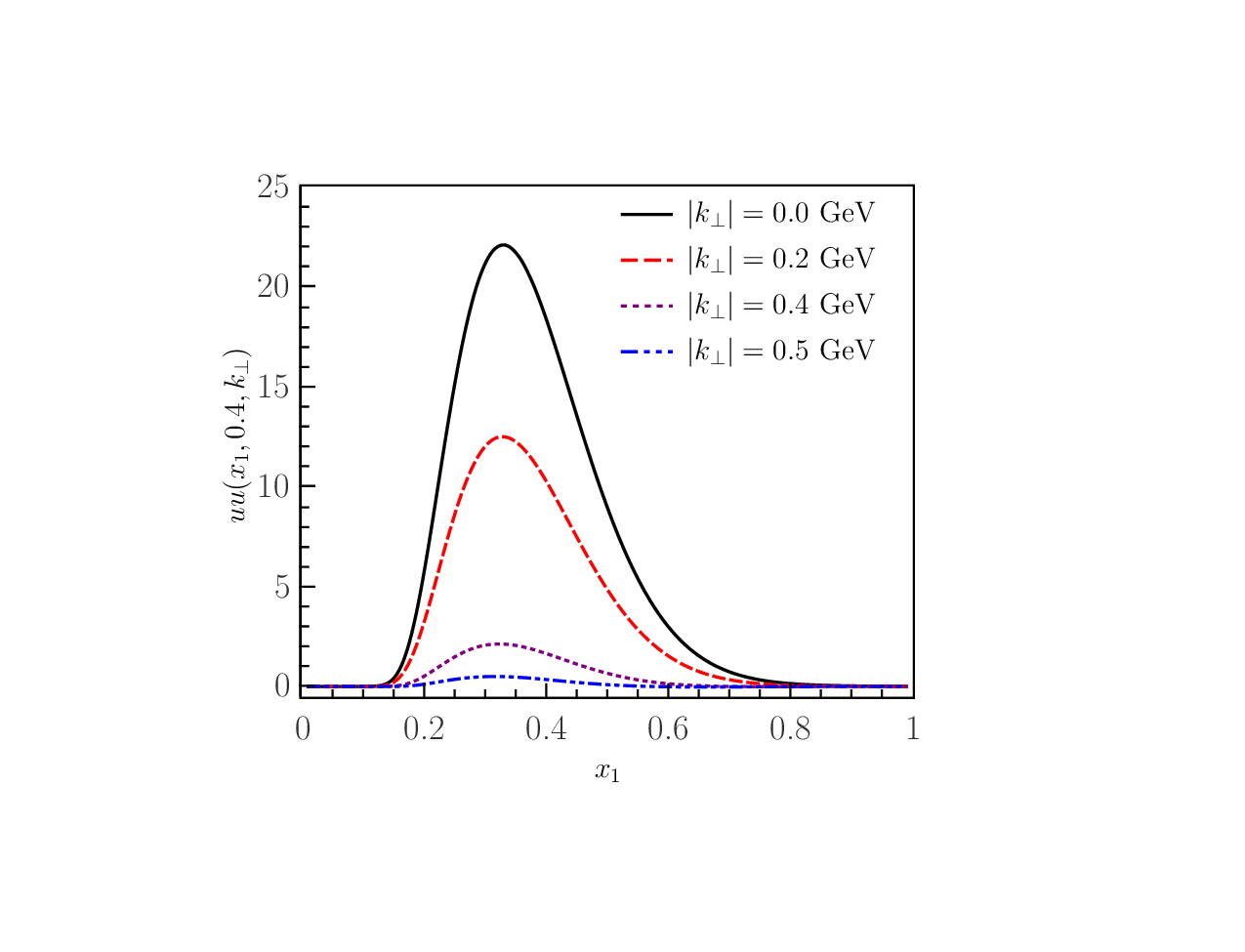}
%\vskip -2.5cm
\caption{The same as in Fig. 1, but in the IK model,
Eq. (\ref{ik}).}
\thispagestyle{empty}
\end{minipage}
\end{figure}
By inserting the wave function Eq. (\ref{ikwf}) in Eq. (\ref{q1q2}),
one gets instead, in the IK model:
\begin{eqnarray}
{uu}^{IK}(x_1,x_2,k_\perp) =
a_{\cal {S}}^2 \, uu_{\cal {S}} +
a_{\cal {S'}}^2 \, uu_{\cal {S'}} +
a_{\cal {S}}a_{\cal {S'}} \, uu_{\cal {SS'}} +
a_{\cal M}^2 \, uu_{\cal M},
\label{ik}
\end{eqnarray} 
where $uu_{\cal {S}}=uu^{HO}$ and
\begin{eqnarray} 
\label{uus'}
uu_{\cal {S'}} (x_1,x_2,k_\perp) & = & 
{C \over 3} {{e^{ - {  k_\perp^2 \over 2 \alpha^2} }}} 
\int {d \vec k_1 } 
{d \vec k_2 } e^{ - f(\vec k_1, \vec k_2,\alpha) }
\{ [ f(\vec k_1, \vec k_2,\alpha) + k_\perp^2 / ( 2 \alpha^2 ) ]^2
\nonumber
\\
& - & [\vec k_\perp \cdot (\vec k_1 - \vec k_2)/\alpha^2]^2
- 3 [2 f(\vec k_1, \vec k_2,\alpha) + k_\perp^2/\alpha^2] + 9 \}
\nonumber
\\
& \times & 
\delta \left ( x_1 - {k_1^- \over P^-} \right )
\delta \left ( x_2 - {k_2^- \over P^-} \right )~,
\end{eqnarray}
\begin{eqnarray} 
\label{uus's}
uu_{\cal {SS'}} (x_1,x_2,k_\perp) & = & 
{C \over \sqrt{3}} {{e^{ - {  k_\perp^2 \over 2 \alpha^2}}}} 
\int {d \vec k_1 } 
{d \vec k_2 } e^{ - f(\vec k_1, \vec k_2,\alpha)}
\{ 
2 f(\vec k_1, \vec k_2,\alpha) + k_\perp^2/\alpha^2 - 6 \}
\nonumber
\\
& \times & 
\delta \left ( x_1 - {k_1^- \over P^-} \right )
\delta \left ( x_2 - {k_2^- \over P^-} \right )~,
\end{eqnarray}
\begin{eqnarray} 
uu_{\cal {M}} (x_1,x_2, k_\perp) & = & 
{C \over 3 \alpha^4} {{e^{ - {  k_\perp^2 \over 2 \alpha^2}}}} 
\int {d \vec k_1 } 
{d \vec k_2 } e^{ - f(\vec k_1, \vec k_2,\alpha)}
\left \{  \frac{1}{2}  
\left[k_1^2+k_2^2+4\vec{k}_1 \cdot \vec{k}_2-\frac{k^2_{\perp}}{2}\right]^2
- \frac{1}{2} 
\left[ \vec{k}_{\perp} \cdot (\vec{k}_1-\vec{k}_2) \right]^2 \right. 
\nonumber
\\
& + &
\left.
\frac{32}{3} \pi^2 
\left \{ 
(k_1^2-k_2^2)^2- 
\left[\vec{k}_{\perp}
\cdot (\vec{k}_1+\vec{k}_2) \right]^2 
\right \} 
g (\vec k_1, \vec k_2, \vec k_\perp) \right \}
\nonumber
\\
& \times &
\delta \left ( x_1 - {k_1^- \over P^-} \right )
\delta \left ( x_2 - {k_2^- \over P^-} \right )~.
\label{uum}
\end{eqnarray}
where
\begin{eqnarray}
g (\vec k_1, \vec k_2, \vec k_\perp)
& = &
{\underset{m_1}\sum}  \bra 1~m_1~1~-m_1~|~0~0\ket 
 {Y}^{\ast}_{1m_1}(\Omega_{\rho})
 {Y}^{\ast}_{1-m_1}(\Omega_{\lambda})
\nonumber
\\
& \times &
{\underset{m_2}\sum}  \bra 1~m_2~1~-m_2~|~0~0\ket 
{Y}^{}_{1 m_2}(\Omega_{\rho'})
{Y}^{}_{1-m_2}(\Omega_{\lambda}) 
\nonumber
\\
& = &
\frac{3}{16\pi^2}
\left[
\cos{\theta_{{\rho'}}}
\cos{\theta_{{\lambda}}} +
\sin{\theta_{{\rho'}}}
\sin{\theta_{{\lambda}}}
\cos{(\phi_{{\lambda}}-\phi_{{\rho'}})}
\right]
\nonumber
\\
& \times &
\left[
\cos{\theta_{{\rho}}}
\cos{\theta_{{\lambda}}}+
\sin{\theta_{{\rho}}}
\sin{\theta_{{\lambda}}}
\cos{(\phi_{{\lambda}}-\phi_{{\rho}})}
\right]~.
\label{gk}
\end{eqnarray}
{  In Eq. (\ref{gk}),
the solid angles $\Omega_\lambda$, $\Omega_\rho$, $\Omega_{\rho'}$
are the ones defined by the vectors
$\vec{k}_{\rho'} = ( \vec{k}_1 - \vec{k}_2 + 
\vec{k}_{\perp} ) / \sqrt{2} $,
$\vec{k}_{\rho} = ({\vec{k}_1-\vec{k}_2-\vec{k}_{\perp}})/ \sqrt{2} $ and 
$\vec{k}_{\lambda}= 
\sqrt{3}(\vec{k}_1+\vec{k}_2)/{\sqrt{2}}$, so that
the angles $\theta_{\rho(\rho')},\phi_{\rho(\rho')},
\theta_\lambda,\phi_\lambda$ appearing in Eq. (\ref{gk})
can be written therefore
through the components of $\vec k_1$, $\vec k_2$, $\vec k_\perp$
according to the transformation laws Eq. (\ref{coor}).

{  It is clear from 
Eqs. (\ref{uus'}-\ref{gk}) that in the IK
model there is an additional dependence on $k_\perp$,
aside from the Gaussian dependence found in the HO case.}
In particular, in Eq. (\ref{uum}), a peculiar $k_\perp$
dependence is introduced by the presence of 
orbital angular momentum of the quarks.
Anyway, since the correlations under scrutiny are for 
unpolarized protons, no dependence on the orientation of 
$\vec k_\perp$ remains. 
One can easily realize this result mathematically by rewriting 
the above expressions in terms of scalar products. }

\section{Results and discussion}

\begin{figure}[t]
\begin{minipage}[t] {70 mm}
\vspace{7.0cm}
\includegraphics{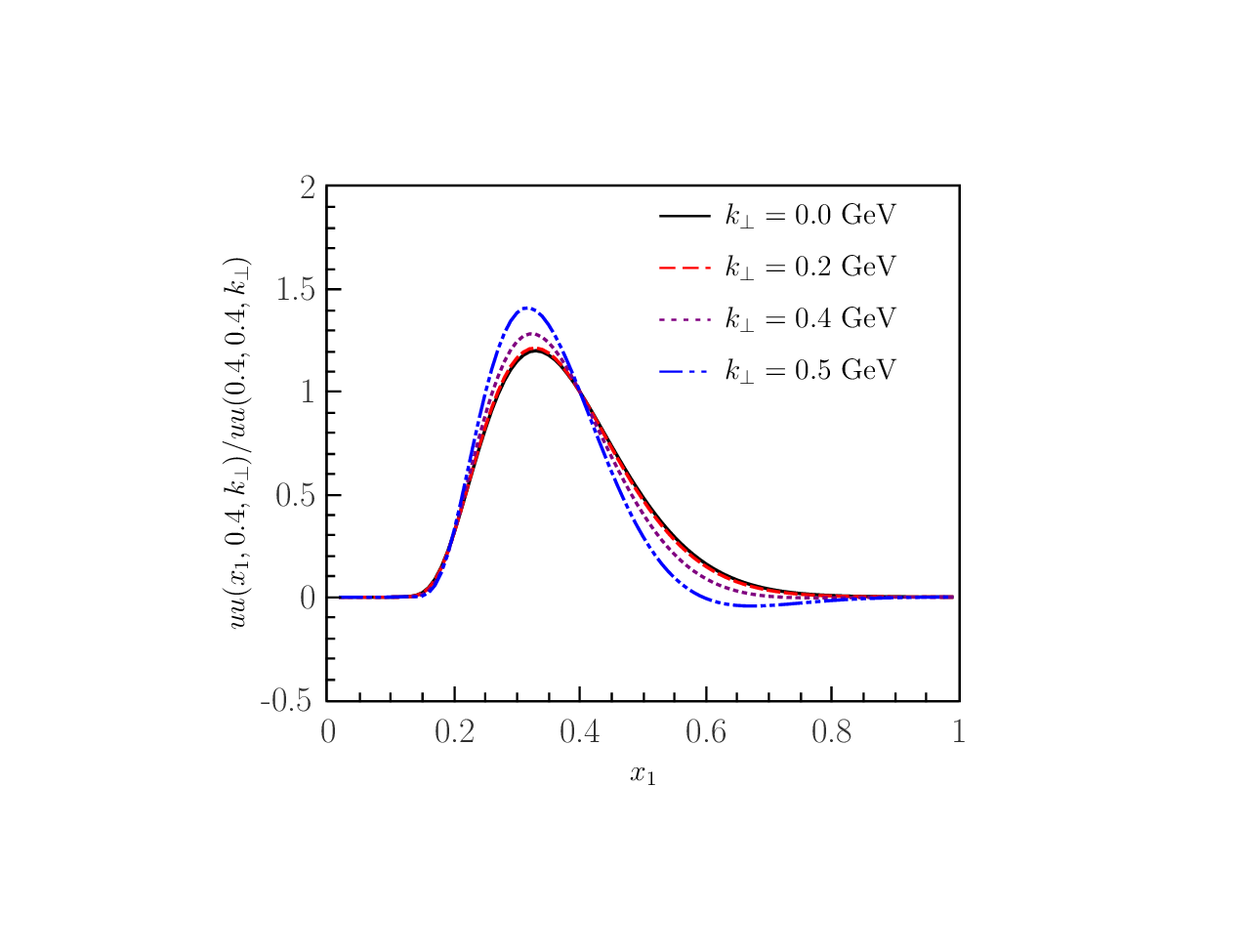}
%\vskip 6.5cm
\caption{The ratio Eq. (\ref{rat1}) in the IK model, for
four values of $k_\perp$.}
\end{minipage}
\hspace{\fill}
\begin{minipage}[t] {70 mm}
\vspace{7.0cm}
\includegraphics{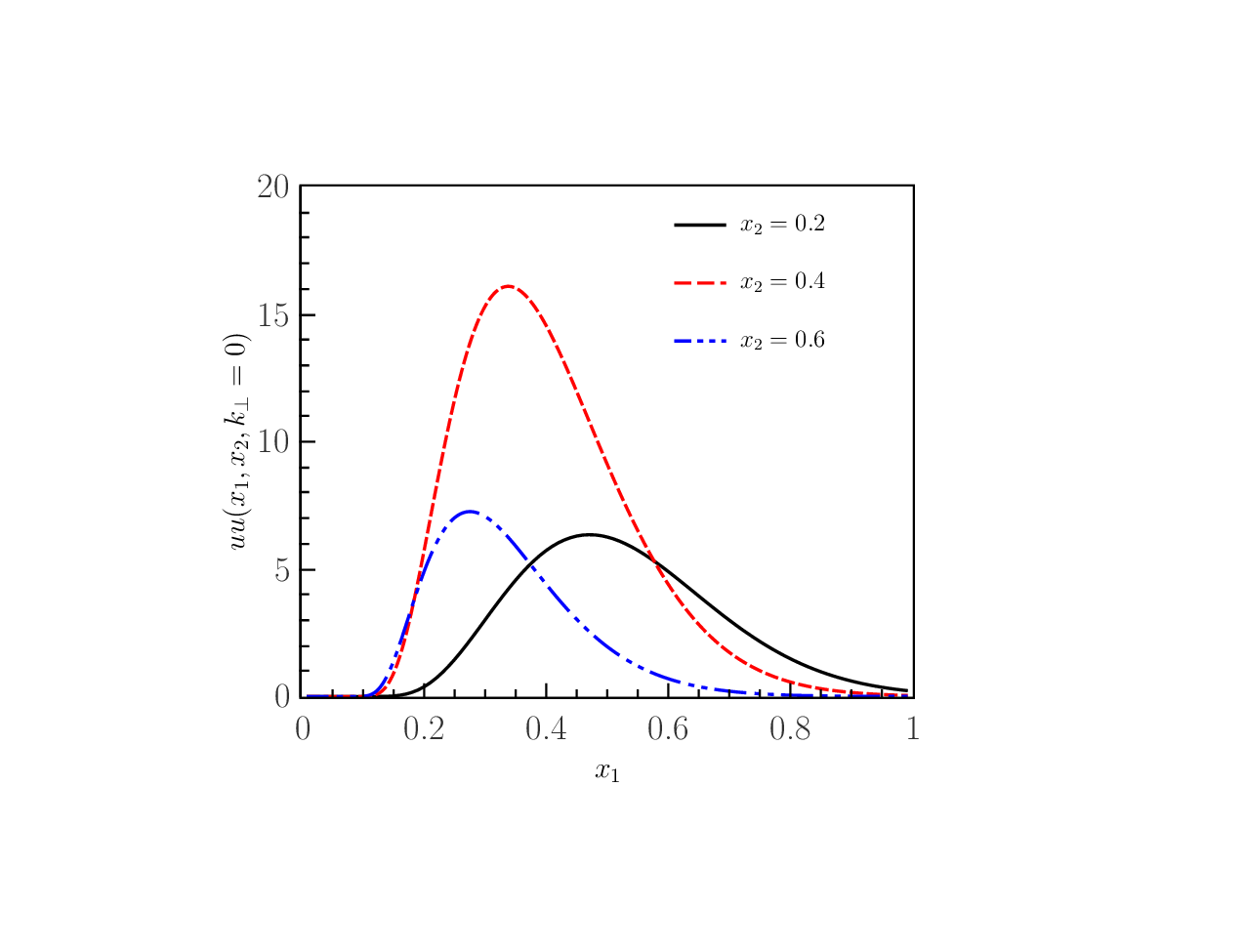}
%\vskip 6.5cm
\caption{The function $uu(x_1,x_2,k_\perp=0)$ 
evaluated in the HO model for three different values of $x_2$.}
\end{minipage}
\end{figure}

The results of the calculation are collected in Figs. 1-7.
We have decided to show basically the same quantities presented
in Ref. \cite{man_bag}, because they describe all the features we
want to emphasize and moreover they allow
an easy comparison of the results of the
different models. 

The dPDF $uu(x_1,x_2,k_\perp)$ evaluated in the HO model,
Eq. (\ref{uuho}), for $x_2=0.4$ at four values of $k_\perp$,
is shown in Fig. 1.
These results are
qualitatively and quantitatively similar to those
in Ref. \cite{man_bag} once correlations are added
to the simplest bag model scheme. As already stated
several times, in the present scheme correlations are
present from the very beginning.
The $k_\perp$ dependence found here is Gaussian,
as a consequence of the HO wave functions used.
Also in the model of Ref. \cite{man_bag} a similar Gaussian behavior
was found.
In Fig. 2 , the same is shown in the IK model, using
Eq. (\ref{ik}).
Although the results are qualitatively similar and 
even quantitatively not far
from the HO ones, relevant differences are found, as
illustrated in the next two figures.
As a matter of facts,
in the HO calculation, the approximation Eq. (\ref{eq:zfact})
perfectly holds:
the $k_\perp$ dependence is only in the
exponent of the Gaussian function outside the integral
in Eq. (\ref{uuho}). 
In the IK model, it is not the case, as it is seen in Figs. 3,
where the following ratio is shown
\begin{equation}
r_1(x_1,k_\perp) = { uu(x_1,x_2=0.4,k_\perp)
\over
uu(x_1=0.4,x_2=0.4, k_\perp)}~,
\label{rat1}
\end{equation}
which should be $k_\perp$ independent if
the approximation
Eq. (\ref{eq:zfact}) were correct.
Actually, a {  small} violation, comparable in size to the one found in Ref.
\cite{man_bag} 
%once the simple bag scheme is properly modified,
is obtained. 
In the present calculation it is perfectly clear that the size
of the violation of the approximation Eq. (\ref{eq:zfact}) is model
dependent: it holds exactly if the ground state of the 
HO is taken, it is violated if mixture of states, with given
OAM, are added.
The fact that it is violated weakly is, again, a model dependent
feature. In a model with stronger SU(6) breaking
with respect to the IK one,
strong violations, related to the OAM content
of the proton, would arise.

\begin{figure}[t]
\begin{minipage}[t] {70 mm}
\vspace{7.0cm}
\includegraphics{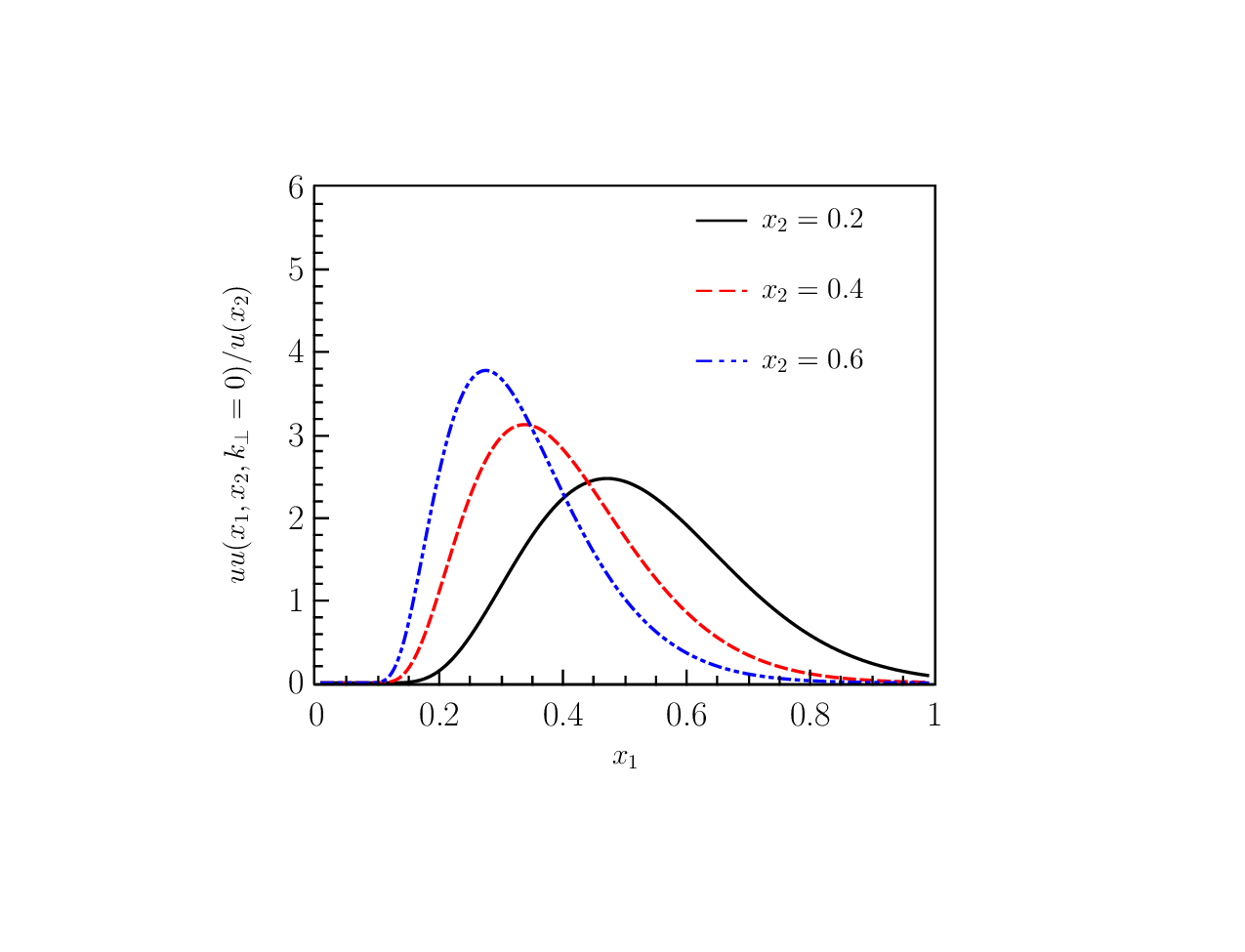}
%\vskip -2.5cm
\caption{The ratio Eq. (\ref{rat2}) in the HO model, for three
different values of $x_2$.}
\thispagestyle{empty}
\end{minipage}
\hspace{\fill}
\begin{minipage}[t] {70 mm}
\vspace{7.0cm}
\includegraphics{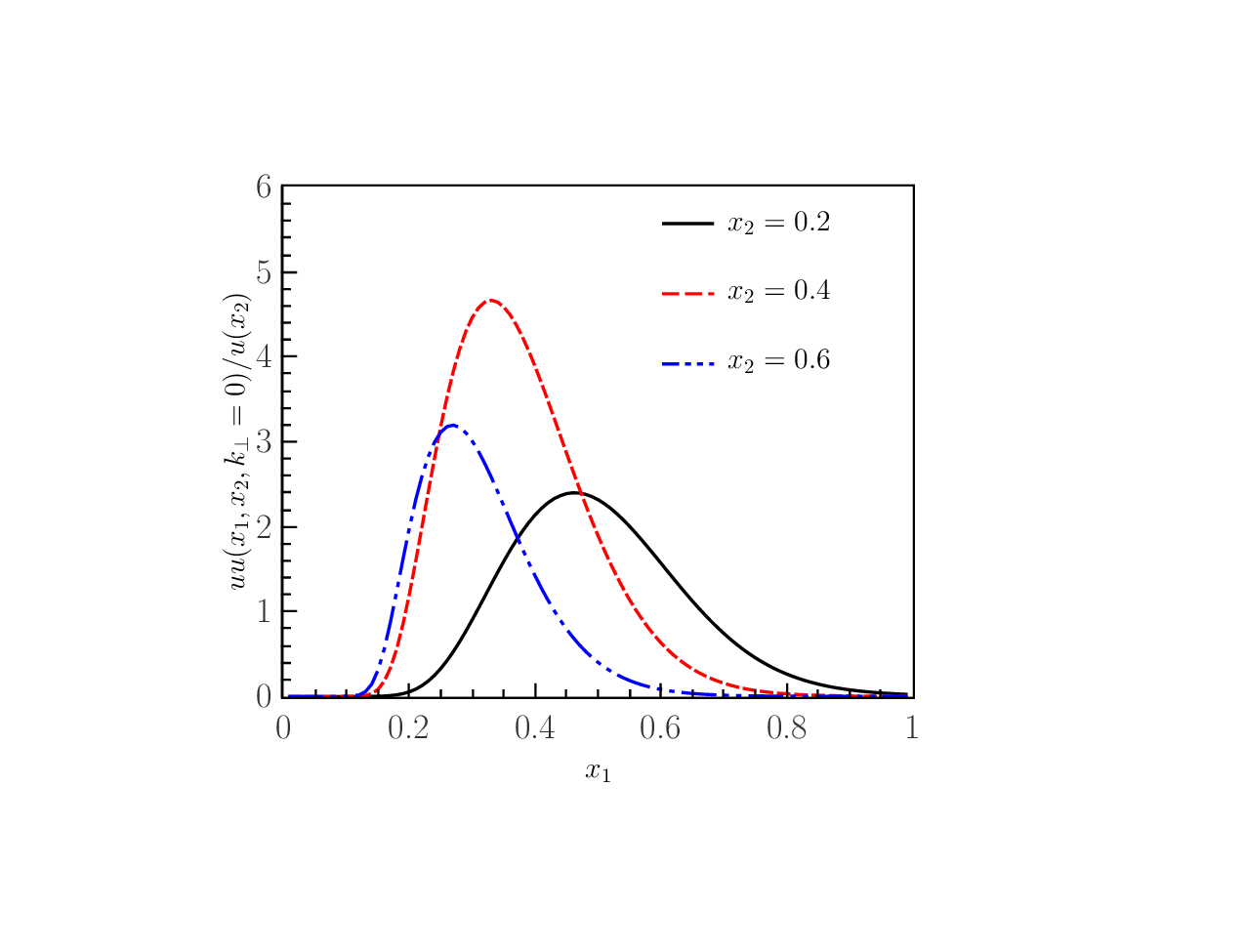}
%\vskip -2.5cm
\caption{The same as in Fig. 5, but in the IK model.}
\thispagestyle{empty}
\end{minipage}
\end{figure}

The extent to which the other approximation, Eq. (\ref{eq:xfact}), is
violated, is shown in Figs. 4-7. 
In particular, the ratio 
\begin{equation}
r_2(x_1,x_2) = { uu(x_1,x_2, k_\perp=0) \over u(x_2) }~,
\label{rat2}
\end{equation}
where $u(x_2)$ is the standard PDF, 
shown in Fig. 5 (6) in the HO (IK) model,
should not depend on the choice of $x_2$, if the
approximation Eq. (\ref{eq:xfact}) were valid.
Results in Fig. 4 are qualitatively similar
to the ones in Ref. \cite{man_bag}: for example the maxima
move towards lower $x_1$ values by increasing $x_2$.
In Fig. 5, in which $r_2$ is
plotted for the HO model,
this trend is confirmed but the maxima become bigger when
$x_2$ gets bigger, a behavior which is
opposite to the one found in the corresponding figure of Ref.
\cite{man_bag}. 
{In Fig. 6, in which $r_2$ is
plotted for the 
IK model, the trend
of the HO model is basically confirmed, although
the curve at $x_2 = 0.4$ has a peculiar behavior,
confirming the model dependence of this ratio.
This can be understood thinking to the different shape
of the single particle PDFs appearing in the denominator.
For example, the difference with the bag results is understood
realizing that, in the latter framework, 
PDFs are 
not vanishing at $x=0$ and present
a sizable tail at $x <0$ (see, e.g. Ref. \cite{Jaffe:1974nj}), 
while the same does not happen using constituent quark model, where  
a {  few percent} support violation is found only at $x>1$.}

{  The issue of support violation in the present dPDFs calculation
deserves  further discussion. As a matter of fact, since
dPDFs involve two quarks momenta, $x_1$ and $x_2$,
with $x_1 + x_2 < 1$, in their evaluation the total support violation turns
out to be greater than in the case of the PDFs. For example, looking at Fig. 4,
it is clear that, while the curve corresponding to $x_2=0.6$
should go to zero already at $x_1 = 0.4$, a rather sizable 
tail is obtained, in the HO calculation, in the non-physical region.
The same Figure shows that the situation gets worse
by increasing $x_2$.
Besides, comparing Figs. 1 and 2, the IK model
seems to work slightly better than the HO one.
Quantitatively, the amount of support violation, evaluated as the
fraction of quark momentum in the non-physical region, is, 
at $k_\perp = 0$, in the
HO (IK) model, 5 \% ( 4 \%) at $x_2=0.2$, 9 \% (7 \%) at $x_2 = 0.4$,
18 \% ( 15 \%) at $x_2 = 0.6$. 
The IK model is confirmed to work a little better than the HO one.
However, since a severe support violation could originate serious problems
when pQCD evolution is applied to the model results,
a proper framework to evaluate dPDFs 
could be one where the problem does not arise, like,
e.g., light-front dynamics \cite{LF1,LF2}.
We are presently investigating this scenario.}
\begin{figure}[t]
\vspace{10.7cm}
\includegraphics{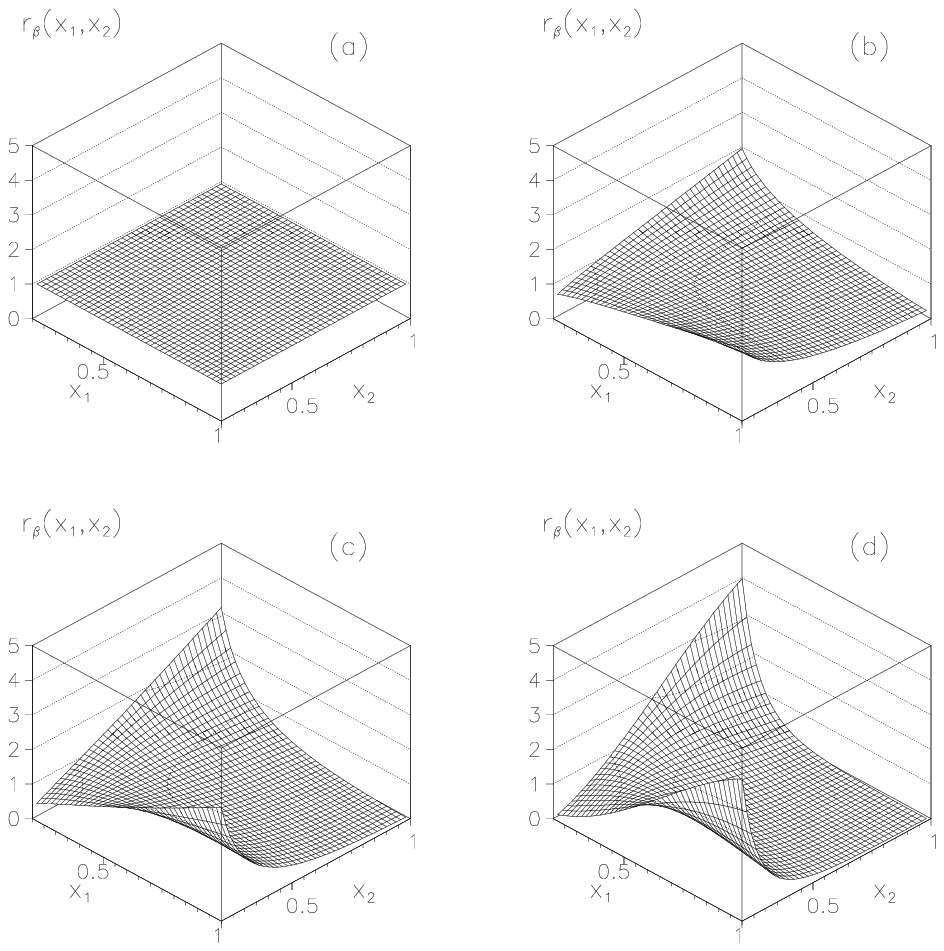}
%\vskip 6.5cm
\caption{The ratio Eq. (\ref{rb}), for:
a) $\beta = 0$, i.e., in an uncorrelated scenario; b) $\beta = 0.25$;
c) $\beta = 0.5$; d) $\beta =1$, i.e., in the correlated HO
framework. {  In the figure, for presentation convenience, 
the variables $x_1$ and $x_2$ range between 0.2 and 1.}}
\end{figure}

The strong violation of Eq. (\ref{eq:xfact}) is also reported in Fig. 6.
This figure has been drawn thinking that the HO framework is
indeed a strongly correlated one.
By defining dPDFs $uu$ and PDFs $u$ introducing a parameter $\beta$,
as follows
\begin{equation}
{uu}_\beta(x_1,x_2,k_\perp=0) = 
2 { (4 - \beta)^{3/2} \over \pi^3 \alpha^6 } 
\int {d \vec k_1 } 
{d \vec k_2 } e^{ 
- 2 (k_1^2 + k_2^2 + \beta \vec k_1 \cdot \vec k_2)/\alpha^2}
\delta \left ( x_1 - {k_1^- \over P^-} \right )
\delta \left ( x_2 - {k_2^- \over P^-} \right )~,
\label{uubeta}
\end{equation}
\begin{equation}
{u}_\beta(x_i) = 
2 { (4 - \beta)^{3/2} \over \pi^{3} \alpha^6 } 
\int {d \vec k_1 } 
{d \vec k_2 } e^{ 
- 2 (k_1^2 + k_2^2 + \beta \vec k_1 \cdot \vec k_2)/\alpha^2}
\delta \left ( x_i - {k_i^- \over P^-} \right )~,
\label{ubeta}
\end{equation}
it is easily seen that the HO model, a very strongly correlated
one due to the HO potential, is recovered for $\beta=1$.
$\beta = 0$ represents instead an uncorrelated model, like, e.g.,
the genuine bag in the cavity approximation. 
Intermediate values of $\beta$ may be
similar to the modified bag model shown in Ref. \cite{man_bag}.
This behavior is found indeed, as demonstrated by Fig. 7, where
the ratio
\begin{equation}
r_\beta(x_1,x_2) = { 2 uu_\beta(x_1,x_2,k_\perp=0)
\over u_\beta(x_1) u_\beta(x_2) }
\label{rb}
\end{equation}
is shown.
The ratio we obtain is actually rather flat 
{  and approximately zero for $x_1\simeq x_2\simeq 1$,
i.e. in the extreme non-physical region, as expected. 
In the same region, due possibly to momentum non-conservation
in the bag model, some structures are seen in the calculation 
of Ref. \cite{man_bag}. }

{We remind that  
%A comment is in order concerning the normalization of 
%the $uu$ dPDF and the $u$ PDF. As it is seen from Eq. (\ref{ubeta}),
the distribution $u$ has norm 2, i.e., the same of
$uu$ (cf. Eqs. (\ref{norm}) and (\ref{ubeta})).
This is in agreement with the number density interpretation
of both distributions, and with the number sum rule in the valence sector
derived in Ref. \cite{Gaunt:2009re}.
One should notice that, for presentation convenience, in the numerator
of Eq. (\ref{rb}) the $uu$ distribution has been multiplied by 
a factor of two. In this way, a ratio equal to one 
is obtained when the $x_1,x_2$ factorization holds.}
 
In closing this section, an important caveat is in order.
We reiterate that the present estimate of quark correlations,
as the one of Ref. \cite{man_bag}, showing important violations
of the often assumed approximations Eqs. (\ref{eq:zfact}) and (\ref{eq:xfact}),
is expected to be reliable only in the valence quark region, i.e.
for $x>0.1$, and at the low energy scale described by the CQM.
Actual data from LHC are dominated by the very low $x$ region
($x<10^{-3}$), at a very high scale of momentum transfer.
Any comparison of this kind of results with data is instructive only
once  pQCD effects are implemented by using QCD evolution
to the experimental scale.
Certainly, due to the coupling of quarks
and gluons, correlations will be generated 
even if exact factorization is assumed at the low energy scale of the model.

\section{Conclusions}

In many phenomenological applications, relevant, e.g., for the LHC 
experimental program,
parton correlations are neglected in modelling
double parton distribution functions.
The latter quantities are therefore factorized in two terms, 
one depending on the longitudinal
momentum fractions of the two quarks, $x_1$ and $x_2$, and another 
on the transverse momentum separation, $\vec k_\perp$, e.g. a sort
of $x - \vec k_\perp$ factorization is assumed.
In addition to that, also an $x_1-x_2$ factorization assumption
is usually made, e.g., the distribution
on the longitudinal momenta $x_1, x_2$ is taken to be uncorrelated. 
This study aims at establishing to what extent constituent quark models
support the two types of factorization,
comparing the results
with those of an important previous analysis, 
performed in a properly modified version of the simplest bag model.
In this way, model-dependent features can be
distinguished from model-independent ones.
An analysis of double quark correlations has been therefore performed
in the valence quark region, for spin averaged double parton
distribution functions.
Two different constituent quark models have been used, i.e.,
a simple non relativistic SU(6) symmetric model
and its generalization, the model of Isgur and Karl,
corresponding to a mild SU(6) breaking due to the
exchange of one gluon. 
For the aim of studying correlations and of understanding their
dynamical origin,
the framework used here, correlated from the very
beginning through the interquark potential, is to be preferred
with respect to models where correlations are added
through prescriptions modifying independent particle models,
such as the simplest version of MIT bag model.  
The conclusions of our work are as follows.
In both constituent quark models, 
the $x_1-x_2$ factorization is strongly violated,
confirming a similar conclusion obtained already in the bag model
framework. This feature seems therefore a model-independent one.
In the SU(6) symmetric model, where the proton is described
by a pure symmetric ${\cal S}$ wave, 
exact $x - k_\perp$ factorization is obtained.
This features is mildly spoiled in the Isgur and Karl model,
as it happens in the bag model framework. 
With respect to the latter,
the present approach allows to understand more clearly the dynamical
origin of the breaking of the $x -  k_\perp$ factorization,
{ and its relation to the orbital angular momentum of the quarks}. 
It is also found that
the amount of violation of the $x -  k_\perp$ factorization
is a model dependent feature. Certainly this effect is a small one
in the scenario discussed here, where the symmetric ${\cal S}$ wave
is dominating. Models with more sizable contributions from
components of the wave function with higher values
of the orbital angular momentum would certainly show a stronger violation.
In closing, it is important to stress that, for the results
of model calculations of double parton distributions, obtained either
in a bag model or in a constituent quark one,
to be phenomenologically relevant to the LHC Physics programme,
the evaluation of their QCD evolution is crucial.
This will be the subject of further studies, together with the evaluation
of the same observables using relativistic, Light-Front approaches.
In this framework, successfully applied in Hadronic Physics 
in general (see, e.g., \cite{LF1}) and for the calculation
of PDFs in particular (see, e.g., \cite{LF2}),
important contributions from high values of the relative
orbital angular momentum naturally arise.

\section{Acknowledgments}
{ We thank Markus Diehl for fruitful comments.}
S.S. thanks Livio Fan\`o for useful
discussions.
{  This work was supported in part by the Research Infrastructure
Integrating Activity Study of Strongly Interacting Matter (acronym
HadronPhysic3, Grant Agreement n. 283286 and n. 283288) 
under the Seventh Framework
Programme of the European Community,}
by the Mineco
under contract FPA2010-21750-C02-01,
by GVA-Prometeo/2009/129, and by CPAN(CSD-00042).
S.S. thanks the Department
of Theoretical Physics of the University of Valencia for warm hospitality,
the { IVICFA} and the project ``Partonic structure of
mesons, nucleons and light nuclei'' of the 
INFN-Mineco agreement for
financial support. 
V.V. thanks the INFN, sezione di Perugia and the Department
of Physics of the University of Perugia for warm hospitality
and support.


\begin{thebibliography}{34}%

%\cite{Paver:1982yp}
\bibitem{Paver:1982yp}
  N.~Paver and D.~Treleani,
  %``Multi - Quark Scattering And Large P(t) Jet Production In Hadronic Collisions,''
  Nuovo Cim.\ A {\bf 70} (1982) 215.
  %%CITATION = NUCIA,A70,215;%%
  %135 citations counted in INSPIRE as of 26 Feb 2013
%\cite{Akesson:1986iv}
\bibitem{livio} 
  T.~Akesson {\it et al.}  [Axial Field Spectrometer Collaboration],
  %``DOUBLE PARTON SCATTERING IN p p COLLISIONS AT S**(1/2) = 63-GeV,''
  Z.\ Phys.\ C {\bf 34}, 163 (1987).
  %%CITATION = ZEPYA,C34,163;%%
  %144 citations counted in INSPIRE as of 06 Mar 2013
\bibitem{pg}
%\cite{Bartalini:2010su}
  P.~Bartalini, (ed.) and L.~Fan\`o, (ed.),
``Multiple partonic interactions at the LHC. Proceedings, 1st International 
Workshop, MPI'08, Perugia, Italy, October 27-31, 2008,''
arXiv:1003.4220 [hep-ex].
%\cite{Kulesza:1999zh}
\bibitem{Kulesza:1999zh} 
  A.~Kulesza and W.~J.~Stirling,
  %``Like sign $W$ boson production at the LHC as a probe of double parton scattering,''
  Phys.\ Lett.\ B {\bf 475}, 168 (2000).
%  [hep-ph/9912232].
  %%CITATION = HEP-PH/9912232;%%
  %50 citations counted in INSPIRE as of 26 Feb 2013
%\cite{Cattaruzza:2005nu}
\bibitem{Cattaruzza:2005nu} 
  E.~Cattaruzza, A.~Del Fabbro and D.~Treleani,
  %``Fractional momentum correlations in multiple production of $W$ bosons and of $b \bar{b}$ pairs in high energy $p p$ collisions,''
  Phys.\ Rev.\ D {\bf 72}, 034022 (2005).
%  [hep-ph/0507052].
  %%CITATION = HEP-PH/0507052;%%
  %25 citations counted in INSPIRE as of 26 Feb 2013
%\cite{Maina:2009sj}
\bibitem{Maina:2009sj} 
  E.~Maina,
  %``Multiple Parton Interactions in Z+4j, W+- W+- + 0/2j and W+ W- + 2j production at the LHC,''
  JHEP {\bf 0909}, 081 (2009).
%  [arXiv:0909.1586 [hep-ph]].
  %%CITATION = ARXIV:0909.1586;%%
  %39 citations counted in INSPIRE as of 26 Feb 2013
%\cite{Gaunt:2010pi}
\bibitem{Gaunt:2010pi} 
  J.~R.~Gaunt, C.~-H.~Kom, A.~Kulesza and W.~J.~Stirling,
  %``Same-sign W pair production as a probe of double parton scattering at the LHC,''
  Eur.\ Phys.\ J.\ C {\bf 69}, 53 (2010).
%  [arXiv:1003.3953 [hep-ph]].
  %%CITATION = ARXIV:1003.3953;%%
  %50 citations counted in INSPIRE as of 26 Feb 2013
%\cite{DelFabbro:1999tf}
\bibitem{DelFabbro:1999tf} 
  A.~Del Fabbro and D.~Treleani,
  %``A Double parton scattering background to Higgs boson production at the LHC,''
  Phys.\ Rev.\ D {\bf 61}, 077502 (2000).
%  [hep-ph/9911358].
  %%CITATION = HEP-PH/9911358;%%
  %55 citations counted in INSPIRE as of 26 Feb 2013
%\cite{Hussein:2006xr}
\bibitem{Hussein:2006xr}
  M.~Y.~Hussein,
  %``A Double parton scattering background to associate WH and ZH production at the LHC,''
  Nucl.\ Phys.\ Proc.\ Suppl.\  {\bf 174} (2007) 55.
%  [hep-ph/0610207].
  %%CITATION = HEP-PH/0610207;%%
  %28 citations counted in INSPIRE as of 26 Feb 2013
%\cite{Bandurin:2010gn}
\bibitem{Bandurin:2010gn} 
  D.~Bandurin, G.~Golovanov and N.~Skachkov,
  %``Double parton interactions as a background to associated HW production at the Tevatron,''
  JHEP {\bf 1104}, 054 (2011).
%  [arXiv:1011.2186 [hep-ph]].
  %%CITATION = ARXIV:1011.2186;%%
  %9 citations counted in INSPIRE as of 26 Feb 2013
%\cite{Berger:2011ep}
\bibitem{Berger:2011ep} 
  E.~L.~Berger, C.~B.~Jackson, S.~Quackenbush and G.~Shaughnessy,
  %``Calculation of W b bbar Production via Double Parton Scattering at the LHC,''
  Phys.\ Rev.\ D {\bf 84}, 074021 (2011).
%  [arXiv:1107.3150 [hep-ph]].
  %%CITATION = ARXIV:1107.3150;%%
  %12 citations counted in INSPIRE as of 26 Feb 2013
%\cite{Dobson:2011gka}
%\bibitem{Dobson:2011gka} 
%  E.~Dobson [for the ATLAS and ATLAS Collaborations],
%  %``A measurement of hard double-partonic interactions in $W \rightarrow l\nu$ + 2 jet events using the ATLAS detector at the LHC,''
%  ATLAS-CONF-2011-160.
%  %%CITATION = ATLAS-CONF-2011-160;%%
%\cite{Aad:2013bjm}
{\bibitem{Aad:2013bjm} 
  G.~Aad {\it et al.}  [ATLAS Collaboration],
  %``Measurement of hard double-parton interactions in W->lv+ 2 jet events at sqrt(s)=7 TeV with the ATLAS detector,''
  New J.\ Phys.\  {\bf 15}, 033038 (2013).}
%  [arXiv:1301.6872 [hep-ex]].}
  %%CITATION = ARXIV:1301.6872;%%
  %7 citations counted in INSPIRE as of 22 May 2013
%\cite{Diehl:2011yj}
\bibitem{Diehl:2011yj} 
  M.~Diehl, D.~Ostermeier and A.~Schafer,
  %``Elements of a theory for multiparton interactions in QCD,''
  JHEP {\bf 1203}, 089 (2012).
%  [arXiv:1111.0910 [hep-ph]].
  %%CITATION = ARXIV:1111.0910;%%
  %19 citations counted in INSPIRE as of 26 Feb 2013
%\cite{Manohar:2012jr}
\bibitem{Manohar:2012jr} 
  A.~V.~Manohar and W.~J.~Waalewijn,
  %``A QCD Analysis of Double Parton Scattering: Color Correlations, Interference Effects and Evolution,''
  Phys.\ Rev.\ D {\bf 85}, 114009 (2012).
%  [arXiv:1202.3794 [hep-ph]].
  %%CITATION = ARXIV:1202.3794;%%
  %11 citations counted in INSPIRE as of 26 Feb 2013
%\cite{Kasemets:2012pr}
\bibitem{Kasemets:2012pr} 
  T.~Kasemets and M.~Diehl,
  %``Angular correlations in the double Drell-Yan process,''
  JHEP {\bf 1301}, 121 (2013).
%  [arXiv:1210.5434 [hep-ph]].
  %%CITATION = ARXIV:1210.5434;%%
%\cite{Mekhfi:1985dv}
%\cite{Schweitzer:2012hh}
\bibitem{weiss} 
  P.~Schweitzer, M.~Strikman and C.~Weiss,
  %``Intrinsic transverse momentum and parton correlations from dynamical chiral symmetry breaking,''
  JHEP {\bf 1301}, 163 (2013);
%  [arXiv:1210.1267 [hep-ph]].
  %%CITATION = ARXIV:1210.1267;%%
  %1 citations counted in INSPIRE as of 07 Mar 2013
%\cite{Schweitzer:2012dd}
  P.~Schweitzer, M.~Strikman and C.~Weiss,
  %``Intrinsic transverse momentum and parton correlations from nonperturbative short-range interactions,''
  arXiv:1212.4031 [hep-ph].
  %%CITATION = ARXIV:1212.4031;%%
\bibitem{Mekhfi:1985dv} 
  M.~Mekhfi,
  %``Correlations In Color And Spin In Multiparton Processes,''
  Phys.\ Rev.\ D {\bf 32}, 2380 (1985).
  %%CITATION = PHRVA,D32,2380;%%
  %35 citations counted in INSPIRE as of 26 Feb 2013
%\cite{Diehl:2011tt}
\bibitem{Diehl:2011tt} 
  M.~Diehl and A.~Schafer,
  %``Theoretical considerations on multiparton interactions in QCD,''
  Phys.\ Lett.\ B {\bf 698}, 389 (2011).
%  [arXiv:1102.3081 [hep-ph]].
  %%CITATION = ARXIV:1102.3081;%%
  %37 citations counted in INSPIRE as of 26 Feb 2013
%\cite{Kirschner:1979im}
\bibitem{oggi}
%\cite{Diehl:2013mla}
  M.~Diehl and T.~Kasemets,
  %``Positivity bounds on double parton distributions,''
  arXiv:1303.0842 [hep-ph].
  %%CITATION = ARXIV:1303.0842;%%
\bibitem{Kirschner:1979im} 
  R.~Kirschner,
  %``Generalized Lipatov-altarelli-parisi Equations And Jet Calculus Rules,''
  Phys.\ Lett.\ B {\bf 84}, 266 (1979).
  %%CITATION = PHLTA,B84,266;%%
  %33 citations counted in INSPIRE as of 26 Feb 2013
%\cite{Shelest:1982dg}
\bibitem{Shelest:1982dg} 
  V.~P.~Shelest, A.~M.~Snigirev and G.~M.~Zinovev,
  %``The Multiparton Distribution Equations In Qcd,''
  Phys.\ Lett.\ B {\bf 113}, 325 (1982).
  %%CITATION = PHLTA,B113,325;%%
  %34 citations counted in INSPIRE as of 26 Feb 2013
%\cite{Manohar:2012pe}

{
\bibitem{agg1} 
  %\cite{Ryskin:2011kk}
%\bibitem{Ryskin:2011kk} 
  M.~G.~Ryskin and A.~M.~Snigirev,
  %``A Fresh look at double parton scattering,''
  Phys.\ Rev.\ D {\bf 83}, 114047 (2011).
%  [arXiv:1103.3495 [hep-ph]].
  %%CITATION = ARXIV:1103.3495;%%
  %24 citations counted in INSPIRE as of 08 May 2013mla


\bibitem{agg2} 
%\cite{Gaunt:2011cfa}
%\bibitem{Gaunt:2011cfa} 
  J.~R.~Gaunt,
  %``Double parton scattering singularity in one-loop integrals,''
  PoS RADCOR {\bf 2011}, 040 (2011).
  
  % arXiv:1103.1888 [hep-ph]
  %%CITATION = POSCI,RADCOR2011,040;%%

\bibitem{agg3} 
%\cite{Gaunt:2012dd}
%\bibitem{Gaunt:2012dd} 
  J.~R.~Gaunt,
  %``Single Perturbative Splitting Diagrams in Double Parton Scattering,''
  JHEP {\bf 1301}, 042 (2013).
%  [arXiv:1207.0480 [hep-ph]].
  %%CITATION = ARXIV:1207.0480;%%
  %2 citations counted in INSPIRE as of 08 May 2013  


\bibitem{agg4} 
  %\cite{Blok:2011bu}
%\bibitem{Blok:2011bu} 
  B.~Blok, Y.~Dokshitser, L.~Frankfurt and M.~Strikman,
  %``pQCD physics of multiparton interactions,''
  Eur.\ Phys.\ J.\ C {\bf 72}, 1963 (2012).
%  [arXiv:1106.5533 [hep-ph]].
  %%CITATION = ARXIV:1106.5533;%%
  %22 citations counted in INSPIRE as of 08 May 2013mla

} 

%%%%%%%%  
\bibitem{Manohar:DPS2} 
  A.~V.~Manohar and W.~J.~Waalewijn,
  %``What is Double Parton Scattering?,''
  Phys.\ Lett.\ B {\bf 713}, 196 (2012).
%  [arXiv:1202.5034 [hep-ph]].
  %%CITATION = ARXIV:1202.5034;%%
  %9 citations counted in INSPIRE as of 26 Feb 2013
\bibitem{man_bag}
%\cite{Chang:2012nw}
  H.~-M.~Chang, A.~V.~Manohar and W.~J.~Waalewijn,
  %``Double Parton Correlations in the Bag Model,''
  Phys.\  Rev.\  D 87, {\bf 034009} (2013).
%  [arXiv:1211.3132 [hep-ph]].
%%CITATION = ARXIV:1211.3132;%%
%\cite{Chodos:1974pn}
\bibitem{Chodos:1974pn} 
  A.~Chodos, R.~L.~Jaffe, K.~Johnson and C.~B.~Thorn,
  %``Baryon Structure in the Bag Theory,''
  Phys.\ Rev.\ D {\bf 10}, 2599 (1974).
  %%CITATION = PHRVA,D10,2599;%%
  %877 citations counted in INSPIRE as of 26 Feb 2013
%\cite{Jaffe:1974nj}
\bibitem{Jaffe:1974nj} 
  R.~L.~Jaffe,
  %``Deep Inelastic Structure Functions in an Approximation to the Bag Theory,''
  Phys.\ Rev.\ D {\bf 11}, 1953 (1975).
  %%CITATION = PHRVA,D11,1953;%%
  %173 citations counted in INSPIRE as of 26 Feb 2013
%\cite{Benesh:1987ie}
\bibitem{Benesh:1987ie} 
  C.~J.~Benesh and G.~A.~Miller,
  %``Deep Inelastic Structure Functions In The Mit Bag Model,''
  Phys.\ Rev.\ D {\bf 36}, 1344 (1987).
  %%CITATION = PHRVA,D36,1344;%%
  %58 citations counted in INSPIRE as of 26 Feb 2013
%\cite{Schreiber:1991tc}
\bibitem{Schreiber:1991tc} 
  A.~W.~Schreiber, A.~I.~Signal and A.~W.~Thomas,
  %``Structure functions in the bag model,''
  Phys.\ Rev.\ D {\bf 44}, 2653 (1991).
  %%CITATION = PHRVA,D44,2653;%%
  %134 citations counted in INSPIRE as of 26 Feb 2013
%\cite{Wang:1982tz}
\bibitem{Wang:1982tz} 
  X.~-M.~Wang, X.~-T.~Song and P.~-C.~Yin,
  %``The Hadron Structure Functions In The Bag Model And Their Modifications,''
  Hadronic J.\  {\bf 6}, 985 (1983).
  %%CITATION = HADJM,6,985;%%
  %7 citations counted in INSPIRE as of 26 Feb 2013
%\cite{Gaunt:2009re}
%\cite{Korotkikh:2004bz}
\bibitem{snig1} 
  V.~L.~Korotkikh and A.~M.~Snigirev,
  %``Double parton correlations versus factorized distributions,''
  Phys.\ Lett.\ B {\bf 594}, 171 (2004).
%  [hep-ph/0404155].
  %%CITATION = HEP-PH/0404155;%%
  %34 citations counted in INSPIRE as of 06 Mar 2013
\bibitem{Gaunt:2009re} 
  J.~R.~Gaunt and W.~J.~Stirling,
  %``Double Parton Distributions Incorporating Perturbative QCD Evolution and Momentum and Quark Number Sum Rules,''
  JHEP {\bf 1003}, 005 (2010).
%  [arXiv:0910.4347 [hep-ph]].
  %%CITATION = ARXIV:0910.4347;%%
  %53 citations counted in INSPIRE as of 26 Feb 2013
%\cite{Snigirev:2010ds}
\bibitem{snig2} 
  A.~M.~Snigirev,
  %``Asymptotic behavior of double parton distribution functions,''
  Phys.\ Rev.\ D {\bf 83}, 034028 (2011).
%  [arXiv:1010.4874 [hep-ph]].
  %%CITATION = ARXIV:1010.4874;%%
  %8 citations counted in INSPIRE as of 06 Mar 2013
\bibitem{trvlc}
  M.~Traini, A.~Mair, A.~Zambarda and V.~Vento,
  %``Constituent quarks and parton distributions,''
  Nucl.\ Phys.\  A {\bf 614}, 472 (1997).
  %%CITATION = NUPHA,A614,472;%%
\bibitem{h1}
  S.~Scopetta and V.~Vento,
  %``A quark model analysis of the transversity distribution,''
  Phys.\ Lett.\  B {\bf 424}, 25 (1998).
%  [arXiv:hep-ph/9706413].
  %%CITATION = PHLTA,B424,25;%%
  S.~Scopetta and V.~Vento,
  %``A quark model analysis of orbital angular momentum,''
  Phys.\ Lett.\  B {\bf 460}, 8 (1999)
  [Erratum-ibid.\  B {\bf 474}, 235 (2000)].
%  [arXiv:hep-ph/9901324].
  %%CITATION = PHLTA,B460,8;%%
\bibitem{vlc_pg}
  S.~Scopetta and V.~Vento,
  %``Generalized parton distributions in constituent quark models,''
  Eur.\ Phys.\ J.\  A {\bf 16}, 527 (2003).
%  [arXiv:hep-ph/0201265].
  %%CITATION = EPHJA,A16,527;%%
\bibitem{siv_bm}
 %\cite{Courtoy:2008vi}
  A.~Courtoy, F.~Fratini, S.~Scopetta and V.~Vento,
  %``A quark model analysis of the Sivers function,''
  Phys.\ Rev.\  D {\bf 78} (2008) 034002;
%  [arXiv:0801.4347 [hep-ph]].
  %%CITATION = PHRVA,D78,034002;%%
  %\cite{Courtoy:2008dn}
  A.~Courtoy, S.~Scopetta and V.~Vento,
 % ``Model calculations of the Sivers function satisfying the Burkardt Sum
 % Rule,''
  Phys.\ Rev.\  D {\bf 79}, 074001 (2009);
%  [arXiv:0811.1191 [hep-ph]].
  %%CITATION = PHRVA,D79,074001;%%
%\cite{Courtoy:2009pc}
%  A.~Courtoy, S.~Scopetta and V.~Vento,
  %``Analyzing the Boer-Mulders function within different quark models,''
  Phys.\ Rev.\ D {\bf 80}, 074032 (2009).
%  [arXiv:0909.1404 [hep-ph]].
  %%CITATION = ARXIV:0909.1404;%%
  %24 citations counted in INSPIRE as of 13 Feb 2013
%\cite{Mekhfi:1988kj}
\bibitem{Mekhfi:1988kj} 
  M.~Mekhfi and X.~Artru,
  %``Sudakov Suppression Of Color Correlations In Multiparton Scattering,''
  Phys.\ Rev.\ D {\bf 37}, 2618 (1988).
  %%CITATION = PHRVA,D37,2618;%%
  %10 citations counted in INSPIRE as of 26 Feb 2013
\bibitem{ik}
  N.~Isgur and G.~Karl,
  %``P Wave Baryons In The Quark Model,''
  Phys.\ Rev.\  D {\bf 18}, 4187 (1978);
  %%CITATION = PHRVA,D18,4187;%%
 %``Positive Parity Excited Baryons In A Quark Model With Hyperfine
  %Interactions,''
  Phys.\ Rev.\  D {\bf 19}, 2653 (1979)
  [Erratum-ibid.\  D {\bf 23}, 817 (1981)].
  %%CITATION = PHRVA,D19,2653;%%

\bibitem{ruju}
  A.~De R\`ujula, H.~Georgi and S.~L.~Glashow,
  %``Hadron Masses In A Gauge Theory,''
  Phys.\ Rev.\  D {\bf 12}, 147 (1975).
  %%CITATION = PHRVA,D12,147;%%

\bibitem{mmg}
  M.~M.~Giannini,
  %``Electromagnetic excitations in the constituent quark model,''
  Rept.\ Prog.\ Phys.\  {\bf 54}, 453 (1990).
  %%CITATION = RPPHA,54,453;%%
  
\bibitem{ik2}
  N.~Isgur, G.~Karl and R.~Koniuk,
  %``Violations Of SU(6) Selection Rules From Quark Hyperfine Interactions,''
  Phys.\ Rev.\ Lett.\  {\bf 41}, 1269 (1978)
  [Erratum-ibid.\  {\bf 45}, 1738 (1980)];
  %%CITATION = PRLTA,41,1269;%%
 N.~Isgur, G.~Karl and J.~Soffer,
  %``ZEROS IN THE NUCLEON FORM-FACTORS AND THE QUARK MODEL,''
  Phys.\ Rev.\  D {\bf 35}, 1665 (1987).
  %%CITATION = PHRVA,D35,1665;%%

%\cite{Conci:1990kt}
%\bibitem{Conci:1990kt}
%  L.~Conci and M.~Traini,
%  %``Quark Momentum Distribution In Nucleons,''
% Few-Body Systems {\bf 8},  123 (1990).
%  %%CITATION = APASA,8,123;%%

%\cite{Cardarelli:1995dc}
\bibitem{LF1} 
  F.~Cardarelli, E.~Pace, G.~Salm\`e and S.~Simula,
  %``Nucleon and pion electromagnetic form-factors in a light front constituent quark model,''
  Phys.\ Lett.\ B {\bf 357}, 267 (1995).
%  [nucl-th/9507037].
  %%CITATION = NUCL-TH/9507037;%%

\bibitem{LF2} 
%\cite{Faccioli:1998aq}
  P.~Faccioli, M.~Traini and V.~Vento,
  %``Polarized parton distributions and light front dynamics,''
  Nucl.\ Phys.\ A {\bf 656} (1999) 400.
%  [hep-ph/9808201].
  %%CITATION = HEP-PH/9808201;%%

\end{thebibliography}
\end{document}